# Neuronal Circuit Policies

Mathias Lechner [* 1]  Ramin M. Hasani [* 1]  Radu Grosu [1]

## Abstract

We propose an effective way to create interpretable control agents, by *re-purposing* the function of a biological neural circuit model, to govern simulated and real world reinforcement learning (RL) test-beds. We model the tap-withdrawal (TW) neural circuit of the nematode, *C. elegans*, a circuit responsible for the worm's reflexive response to external mechanical touch stimulations, and learn its synaptic and neuronal parameters as a policy for controlling basic RL tasks. We also autonomously park a real rover robot on a predefined trajectory, by deploying such neuronal circuit policies learned in a simulated environment. For reconfiguration of the *purpose* of the TW neural circuit, we adopt a search-based RL algorithm. We show that our neuronal policies perform as good as deep neural network policies with the advantage of realizing interpretable dynamics at the cell level.

## 1. Introduction

Through natural evolution, the nervous system of the nematode, *C. elegans*, structured a near optimal wiring diagram (White et al., 1986). Its stereotypic brain composed of 302 neurons connected through approximately 8000 chemical and electrical synapses (Chen et al., 2006). *C. elegans* exhibits distinct behavioral mechanisms to process complex chemical stimulations (Bargmann, 2006), avoid osmotic regions (Culotti & Russell, 1978), sleep (Nichols et al., 2017), show adaptive behavior (Ardiel & Rankin, 2010), perform mechanosensation (Chalfie et al., 1985b), and to control muscles (Wen et al., 2012).

The functions of many neural circuits within its brain have been identified (Wicks & Rankin, 1995; Chalfie et al., 1985a; Li et al., 2012; Nichols et al., 2017). In particular, a neural circuit which is responsible for inducing a forward/backward locomotion reflex when the worm is mechanically exposed to touch stimulus on its body, has been well-characterized (Chalfie et al., 1985a). The circuit is called tap-withdrawal (TW) and it comprises 9 neuron classes which are wired together by means of chemical and electrical synapses. Synaptic polarities (either being excitatory or inhibitory) of the circuit have then been predicted, suggesting that the circuit realizes a competitive behavior between forward and backward reflexes, in presence of touch stimulations (Wicks & Rankin, 1995; Wicks et al., 1996).

Behavior of the tap-withdrawal (TW) reflexive response is substantially similar to the control agent's reaction in some standard control settings such as the impulse response of a controller operating on an *Inverted Pendulum* (Widrow, 1964; Doya, 2000; Russell & Norvig, 2010), a controller acting on driving an under-powered car, to go up on a steep hill, known as the *Mountain Car* (Moore, 1990; Singh & Sutton, 1996), and a controller acting on the navigation of a rover robot that plans to go from point A to B, on a planned trajectory, with two control commands of angular and linear velocity.

We intend to take advantage of the similarity and reconfigure the synaptic and neuronal parameters of a deterministic dynamic model of the TW neural circuit, in each of the mentioned control settings. We use publicly available reinforcement learning toolkits, to evaluate the performance of our neuronal circuit policies. The environments include the inverted pendulum (Schulman et al., 2017), the continuous mountain car of OpenAI Gym[1] and rllab [2], and the Cart-pole of rllab (Duan et al., 2016). In a real robotic setting, We also determine a control task for a rover robot to park autonomously in a specific parking spot, by a learned TW neuronal policy. For all three control challenges, we preserve the near-optimal wiring structure of the TW circuit and adopt a search-based reinforcement learning (RL) algorithm for synaptic parametrization of the network. The approach is named as *neuronal circuit policies*.

Our principle contribution in this work is to demonstrate the performance of a compact neuronal circuit model from the brain of the *C. elegans* worm, as an interpretable continuous time recurrent neural network, in standard control and RL settings. In our experimental evaluations, we demonstrate that our control agent can achieve the performance of the conventional and the state-of-the-art artificial intelligence (AI) control agents, by solving five RL tasks. We show how a learned neuronal circuit policy in a simulated environment, can be transferred to a real robotic environment. We also demonstrate that the function of the neurons in a learned neuronal network is interpretable.

---

*Equal contribution  [1]Cyber Physical Systems, TU Wien, Austria. Correspondence to: Mathias Lechner <mlechner@tuwien.ac.at>, Ramin Hasani <ramin.hasani@tuwien.ac.at>.

[1] https://github.com/openai/gym
[2] https://github.com/rllab/rllab

## 2. Preliminaries

In this section, we first briefly describe the structure and dynamics of the tap-withdrawal neural circuit. We then introduce the mathematical neuron and synapse models utilized to build up the model of the circuit.

### 2.1. Tap-Withdrawal Neural Circuit Revisit

A mechanically exposed stimulus (i.e. tap) to the petri dish in which the worm inhabits, results in the animal's reflexive response in the form of a forward or backward movement. This response has been named as the *tap-withdrawal reflex*, and the circuit identified to underly such behavior is known as the *tap-withdrawal* (TW) neural circuit (Rankin et al., 1990). The circuit is shown in Figure 1. It is composed of four sensory neurons, PVD and PLM (posterior touch sensors), AVM and ALM (anterior touch sensors), four interneuron classes (AVD, PVC, AVA and AVB), and two subgroup of motor neurons which are abstracted as a forward locomotory neurons, FWD, and backward locomotory neurons, REV. Neurons recurrently synapse into each other with excitatory and inhibitory synaptic links. It has been shown that sensory neurons of the TW circuit get activated as a result of an input tap, and transfer the stimulus through the modulatory interneurons PVC and AVD to the command neurons AVA and AVB (Wicks et al., 1996). The TW reflex is then modulated by a competition between these two command neurons, either resulting in a forward scape response (AVB's activation dominates AVA's) or a reversal scape response. Throughout the paper, we illustrate how such recurrent neuronal network can be deployed in standard RL settings. We first sketch how we modeled neurons and synapses to build up the TW circuit.

### 2.2. Neuron Model

Most of the neurons in *C. elegans* are observed to exhibit electrotonic dynamics (Kato et al., 2015), meaning that electric charges spread passively inside a neuron creating graded potentials. This implies that the neurons are non-spiking. Dynamics of the neurons' membrane potential therefore, were modeled by the well-known, deterministic ordinary differential equation (ODE), the *single-compartment mem-*

*brane equation* (Koch & Segev, 1998):

$$C_m \frac{dv_i}{dt} = G_{Leak}\Big(V_{Leak} - v_i(t)\Big) + \sum_{i=1}^{n} I_{in}^{(i)}, \quad (1)$$

where $C_m, G_{Leak}$ and $V_{Leak}$ are parameters of the neuron and $I_{in}^{(i)}$, stands for the external currents to the cell. We adopted Eq. (1) to govern **interneurons**' dynamics.

For interacting with the environment, We introduced sensory and motor neuron models, separately. A **sensory component** consists of two neurons $S_p, S_n$ and a measurable dynamic system variable, $x$. $S_p$ gets activated when $x$ has a positive value, whereas $S_n$ fires when $x$ is negative. Mathematically, the potential of the neurons $S_p$, and $S_n$, as a function of $x$, can be expressed as

$$S_p(x) := \begin{cases} -70mV & \text{if } x \leq 0 \\ -70mV + \frac{50mV}{x_{max}}x & \text{if } 0 < x \leq x_{max} \\ -20mV & \text{if } x > x_{max} \end{cases} \quad (2)$$

$$S_n(x) := \begin{cases} -70mV & \text{if } x \geq 0 \\ -70mV + \frac{50mV}{x_{min}}x & \text{if } 0 > x \geq x_{min} \\ -20mV & \text{if } x < x_{min}. \end{cases} \quad (3)$$

This maps the region $[x_{min}, x_{max}]$ of system variable $x$, to a membrane potential range of $[-70mV, -20mV]$. Note that the potential range is selected to be close to the biophysics of the nerve cells, where the resting potential is usually set around -70 mV and a neuron can be considered to be active when it has a potential around -20 mV (Hasani et al., 2017).

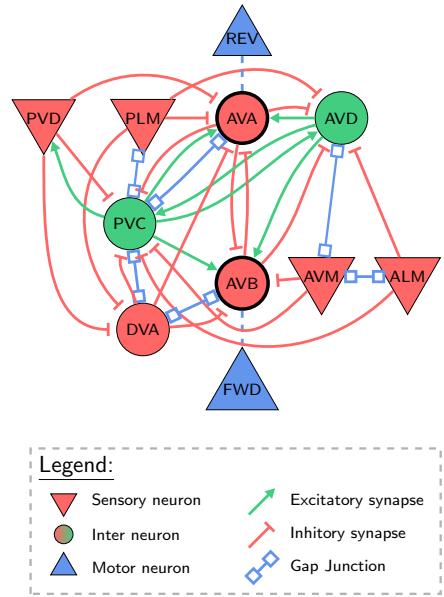

*Figure 1.* Tap-Withdrawal neural circuit schematic.

Similar to sensory neurons, a **motor component** is composed of two neurons $M_n$, $M_p$ and a controllable motor variable $y$. Values of $y$ is computed by $y := y_p + y_n$ and

$$y_p(M_p) := \begin{cases} y_{max}, & \text{if } M_p > -20mV \\ \frac{y_{max}(M_p+70mV)}{50mV}, & \text{if } M_p \in [-70, -20]mV \\ 0, & \text{if } M_p < -70mV \end{cases} \quad (4)$$

$$y_n(M_n) := \begin{cases} y_{min}, & \text{if } M_n > -20mV \\ \frac{y_{min}(M_n+70mV)}{50mV}, & \text{if } M_n \in [-70, -20]mV \\ 0, & \text{if } M_n < -70mV \end{cases} \quad (5)$$

This maps the neuron potentials $M_n$ and $M_p$, to the range $[y_{min}, y_{max}]$. FWD and REV motor classes in Figure 1, are modeled in this fashion.

### 2.3. Synapse Model

**Chemical synapses** are points at which two neurons trade information by the release of neurotransmitters. The chemical synaptic current depends on a non-linear component standing for their conductance strength, which is a function of the presynaptic neurons' potential, $V_{pre}$, and have a maximum weight of $w$, (standing for the maximum conductance of the synapse) as, (Koch & Segev, 1998):

$$g(V_{pre}) = w/1 + e^{\sigma(V_{pre}+\mu)} \quad (6)$$

Moreover, the synaptic current linearly depends on the postsynaptic neuron's membrane potential, $V_{post}$, and therefore can be formulated as, (Koch & Segev, 1998):

$$I_s = g(V_{pre})(E - V_{post}), \quad (7)$$

where by varying $E$, the reversal potential of the synapse, it realizes inhibitory or excitatory connection to their postsynaptic neurons.

An electrical synapse (**gap-junction**), which is a physical junction between two neurons, was modeled by a constant conductance, $\hat{\omega}$, where based on the Ohm's law their bidirectional current between neurons $j$ and $i$, can be computed as

$$\hat{I}_{i,j} = \hat{\omega}\Big(v_j(t) - v_i(t)\Big). \quad (8)$$

For simulating neural networks composed of such dynamic models, we adopted a implicit numerical solver (Press et al., 2007). Formally, we realized the ODE models in a hybrid fashion which combine both implicit and explicit Euler's method. See supplementary materials, Section 2, for a concrete discussion on the model implementation, and the choice of parameters.

Note that one objective of the solver is to be employed in a real-time control system. For reducing the complexity therefore, our method realizes a fixed-step solver. The solver's complexity for each time step $\Delta_t$ is $\mathcal{O}(|\text{\# neurons}|+|\text{\# synapses}|)$. The solver is implemented in C++ in which we construct the TW circuit for performing specific control tasks. We now need to formalize a learning platform to tune the parameters of the circuit for the desired control problem.

### 3. Search-based Reinforcement Learning

In this section we formulate an RL setting for training the parameters of the neural circuit to perform the balancing of the inverted pendulum, control the mountain car, and to park the rover robot.

The behavior of a neural circuit can be expressed as a policy $\pi_\theta(o_i, s_i) \mapsto \langle a_{i+1}, s_{i+1}\rangle$, that maps an observation $o_i$, and an internal state $s_i$ of the circuit, to an action $a_{i+1}$, and a new internal state $s_{i+1}$. This policy acts upon a possible stochastic environment $Env(a_{i+1})$, that provides an observation $o_{i+1}$, and a reward, $r_{i+1}$. The stochastic return is given by $R(\theta) := \sum_{t=1}^{T} r_t$.

Objective of the *Reinforcement learning* is to find a $\theta$ that maximizes $\mathbb{E}\big(R(\theta)\big)$.

Approaches to find such optimal $\theta$, can be categorized into two major groups, based on how randomness is formulated for the environment explorations (Schulman et al., 2015; Salimans et al., 2017): I-Gradient-based and II-search-based methods. The principle of gradient-based RL is to perform random sampling for generating $a_i$, and use the action's influence on the return value, to improve $\theta$ (Williams, 1992). Search-based methods directly randomly sample parame-

---

**Algorithm 1** Random Search + Objective Indicator

**Input:** A stochastic objective indicator $f$,
a starting parameter $\theta$
**Input:** Optimized parameter $\theta$
$f_\theta \leftarrow f(\theta)$
**for** $k \leftarrow 1$ **to** maximum iterations **do**
  $\theta' \leftarrow \theta + rand()$
  $f_{\theta'} \leftarrow f(\theta')$
  **if** $f_{\theta'} < f_\theta$ **then**
    Set $\theta \leftarrow \theta'$
    $f_\theta \leftarrow f_{\theta'}$
    $i \leftarrow 0$
  **end if**
  $i \leftarrow i + 1$
  **if** $i > N$ **then**
    $f_\theta \leftarrow f(\theta)$
  **end if**
**end for**
**return** $\theta$

ters and estimate how good these random parameters are, to update $\theta$ (Salimans et al., 2017; Szita & Lörincz, 2006). Here, we adopted such search-based optimization which can be applied in any control setting, regardless of the internal structure of the policy, (*black-box optimization*). One major obstacle for search-based optimization is the stochastic nature of the RL environment, which makes the objective function, $f(\theta) = \mathbb{E}(R_\theta)$, a probability distribution and the underlying optimization problem an instance of *Stochastic Optimization* (Spall, 2003). Samples of the objective distribution can be generated by running rollouts with $\pi_\theta$ on the environment. A possible solution to overcome a high variance when estimating $\mathbb{E}(R_\theta)$, is to rely on a very large number of samples (Salimans et al., 2017; Duan et al., 2016), which nonetheless comes with high computational costs. Our approach is based on a *Random Search* (RS) (Rastrigin, 1963) optimization, combined with an *Objective Estimate* (OE) as an objective function $f : \theta \mapsto \mathbb{R}^+$.

The OE generates $N$ rollouts with $\pi_\theta$ on the environment and computes an estimate of $\mathbb{E}(R_\theta)$ based on a filtering mechanism on these $N$ samples. We compared two filtering strategies in this context; I-Taking the average of the $N$ samples, and II: taking the average of the worst $k$ samples out of $N$ samples. The first strategy is equivalent to the *Sample Mean estimator*, whereas the second strategy aims to avoid getting mislead by outlying high samples of $\mathbb{E}(R_\theta)$. Our objective for realizing the second strategy was the fact that a suitable parameter $\theta$ makes the policy $\pi_\theta$, control the environment in a decent way even in difficult situations (i.e. rollouts with the lowest return).

In both strategies, to ensure that a single outlying high OE for some $\theta$ does not hinder the algorithm to find a legitimately good parameter, the OE is reevaluated after it is utilized $M$ times, within the algorithm. The full algorithm is outlined in Algorithm 1.

## 4. Experiments

The goal of our experiments is to answer the following questions: 1) How would a neural circuit policy perform in a basic standard control setting? 2) When possible, how would the performance of our learned circuit compare to the other methods? 3) Can we transfer a policy from a simulated environment to a real environment? 4) Can we interpret the behavior of the neural circuit policies?

We prepared five environmental settings for the TW sensory/motor neurons and then deployed our RL algorithm to learn the parameters of the TW circuit to realize the given control objective. Environments include I) Inverted pendulum of Roboschool (Schulman et al., 2017), II) Mountain car of OpenAI Gym, III) Mountain car of rllab, IV) cart-pole balancing of rllab and V) Parking a real rover robot from a learned policy in a simulated environment. The code is available online [1].

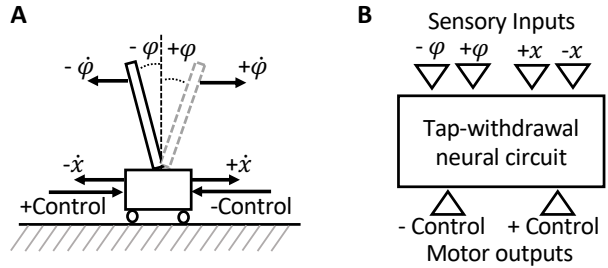

*Figure 2.* Inverted Pendulum setting for the TW circuit. A) observations and control variables in the inverted pendulum balance problem. B) observation and control signals mapping of the pendulum to the TW circuit.

The major constraint that prevents the TW neural circuit (shown in Figure 1) to be tested on arbitrary tasks, is its limited number of sensory and motor neurons. As the TW circuit allows us to incorporate only two input and one output variables, we selected tasks that can be solved with only two observations. We chose environments from different toolkits to evaluate our neural circuit policies on a variety of dynamics, interactions and reward settings. For instance, the two mountain car suites of OpenAI Gym and rllab realize different positive rewards functions as explained in section 4.3.

### 4.1. Inverted pendulum with the TW neural circuit

The TW neural circuit shown in Figure 1, contains four sensory neurons. It therefore, allows us to map the circuit to only two input variables. Note that as we discussed in section 2.2, a sensory component expressed in the form of Eq. (2) and Eq. (3), consists of two neurons for incorporating positive and negative values of the dynamic system variable. The inverted pendulum environment provides four observation variables as shown in Figure 2A. The position of the cart $x$, together with its velocity $\dot{x}$, the angle of the pendulum $\varphi$[2] along with its angular velocity $\dot{\varphi}$.

Since the main objective of the controller is to balance the pendulum in an upward position and make the car stay within the horizontal borders, we fed $\varphi$, and $x$, as the inputs to the sensors of the TW circuit, as illustrated in Figure 2B.

Control commands to the pendulum were originated by the abstract motor neuron classes, FWD and REV components. The activity of these components are governed by Eq. (4)

---

[1]Code for all experiments is available online at: https://github.com/mlech26l/neuronal_circuit_policies

[2]Remark: The environment further splits $\varphi$ into $sin(\varphi)$ and $cos(\varphi)$ to avoid the $2\pi \to 0$ discontinuity

*Table 1.* Mapping the environmental variables to the sensory and motor neurons of the TW circuit, in different experiments

| Experiment | Environment variable | Type | Positive neuron | Negative neuron |
|---|---|---|---|---|
| Inverted Pendulum | $\varphi$ | Sensor (pendulum angle) | PLM | AVM |
|  | $x$ | Sensor (cart position) | ALM | PVD |
|  | $a$ (Control) | Motor (move right/left) | FWD | REV |
| Mountain Car (OpenAI Gym) | $x$ | Sensor (car position) | PLM | AVM |
|  | $\dot{x}$ | Sensor (car's linear velocity) | ALM | PVD |
|  | $a$ (Control) | Motor (move right/left) | FWD | REV |
| Mountain Car (rllab) | $x$ | Sensor (car position) | PLM | AVM |
|  | $\dot{x}$ | Sensor (car's linear velocity) | ALM | PVD |
|  | $a$ (Control) | Motor (move right/left) | FWD | REV |
| Cart-Pole | $\varphi$ | Sensor (pole angle) | PLM | AVM |
|  | $\dot{\varphi}$ | Sensor (pole angular velocity) | ALM | PVD |
|  | $a$ (Control) | Motor (move right/left) | FWD | REV |
| Parking of a Rover | $x$ | Sensor (estimated x position) | PVD |  |
|  | $y$ | Sensor (estimated y position) | PLM |  |
|  | $s$ | Sensor(start signal) | AVM |  |
|  | $\theta$ | Sensor (estimated angular pose) | ALM |  |
|  | $a_1$ (Control) | Motor (angular velocity) | FWD | REV |
|  | $a_2$ (Control) | Motor (linear velocity) | FWD/REV |  |

and Eq. (5), respectively, and represented in Figure 2B, graphically.

We set up the search-based RL algorithm to optimize the neurons and synapses parameters $\omega, \hat{\omega}, \sigma, C_m, V_{Leak}$ and $G_{Leak}$ in the Roboschool `RoboschoolInvertedPendulum-v1` environment with an slight modification (Schulman et al., 2017) in the reward calculation. It is desirable for the cart to stay in the center of the horizontal space. Therefore, we incorporated an additional axillary reward; if the pendulum is in the up-right position and in the center, an additional 20% reward is collected. This bonus linearly decreased. For instance the bonus reward is 10% if the cart is halfway to the end. Right at the borders the bonus reward vanishes. A video of different stages of the learned neuronal circuit policy for the inverted pendulum can be viewed at https://youtu.be/iOHeQ7DhQv8

Note that we were also able the train the TW circuit to solve original version of the inverted pendulum task when feeding $\varphi$, and $\dot{\varphi}$ into the circuit. However, we observed a slow drift of the pendulum toward one of the two ends in final policy, which we wished to eliminate.

### 4.2. The mountain car (OpenAI Gym) control with the TW neural circuit

In this experiment we trained the TW circuit to drive the car shown in Figure 3A uphill to the right-hand side, by generating gravitational momentum. The observation variables are the car's position (on the horizontal axis, $x$) together with its linear velocity. The control signal applies force to the car to move it to the right- and left-hand side, periodically, to finally bring the car up on the hill.

Similar to configuration of the inverted pendulum, we set the observational and motor variables for the tap withdrawal, as illustrated in Figure 3B. The circuit was then learned by the search-based RL algorithm. A video illustrating the control of the car at various episodes during the learning process can be viewed at https://youtu.be/lMrP1sXp3jk.

### 4.3. The mountain car (rllab) control with the TW neural circuit

The environmental observations and motor actions for the mountain car are the same in rllab and OpenAI Gym. Therefore, we set up the TW circuit the same way we did for the mountain car in the Gym.

The major difference of the two environments are in the way the reward is computed. The environment of rllab has a graded continuous reward which is directly associated with the position of the car, whereas in the OpenAI Gym implementation, reward is sparse and gets allocated only once the car reaches a certain altitude. Furthermore, in both frameworks, reward is subjected to a penalty as follows; in rllab the penalty is constant, whereas in the Gym the amount of the penalty varies depending on the amplitude of the performed action. Consequently, energy efficient solutions achieve a higher score, in the Gym environment, in contrast to the rllab version, where the highest scoring solutions bring the car uphill as fast as possible.

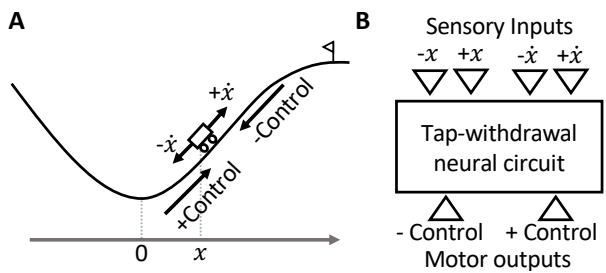

*Figure 3.* The Mountain car setup. A) observations and control variables in the mountain car environment. B) mapping of the observation and control signals to TW circuit.

### 4.4. Cart-pole (rllab) control with the TW circuit

The Cart-pole environmental setting is substantially similar to that of the inverted pendulum. In contrast to the inverted pendulum experiment, here we mapped the observation variables, pole angle, $\varphi$ and the pole's angular velocity $\dot{\varphi}$ to the sensor. As a result, the controller is not aware of the position of the cart, $x$, and whether the boundary of the movable space is nearby or not. An intuitive solution for handling more control variables would be to add sensory neurons to the TW circuit. However, in the present work we intended to maintain the structure of the TW circuit constant, and test its capabilities in control scenarios with the resulting partial observability. Environmental variables for the cart-pole suite were mapped to the circuit elements as denoted in Table 1.

### 4.5. Autonomous parking of the rover with the TW neural circuit

In this experiment, we generalized our TW neuronal circuit policy to a real-world control setting. We let the TW circuit learn to park a rover robot on a determined spot, given a set of checkpoints which form a trajectory, in a deterministic simulated environment. We then deployed the learned policy on a mobile robot in a real environment shown in Figure 4A. The key objective here was to show the capability of the method to perform well in a transformation from a simulated environment to a real setting. For doing this, we developed a *custom deterministic simulated RL environment*[1].

The rover robot provides four observational variables (Starting signal, position of the rover, x, y and its angular pose, $\theta$), together with two motor actions (linear and angular velocity, $v$ and $w$). We mapped all four observatory variables, as illustrated in Figure 4B, to the sensors of the TW. Note that here the geometric reference of the surrounding space is set at the initial position of the robot. Therefore, observation variables are only positive.

We mapped the linear velocity (which is a positive variable throughout the parking task) to one motor neuron and the same variable to another motor neuron. We determined two motor neurons for the positive and negative angular velocity. The mapping details are provided in Table 1. This configu-ration implies that the command neuron, AVA, controls two motor neurons responsible for the turn-right and forward motion-primitives, and AVB to control the turn-left and forward motor neurons. Therefore, the TW circuit is able to govern the robot's locomotion in three different directions, forward, left- and right-turns, which are sufficient to perform a parking trajectory. Furthermore, each command neuron is able to move the rover forward and turn it to the left or right.

[1]https://github.com/mlech26l/neuronal_circuit_policies

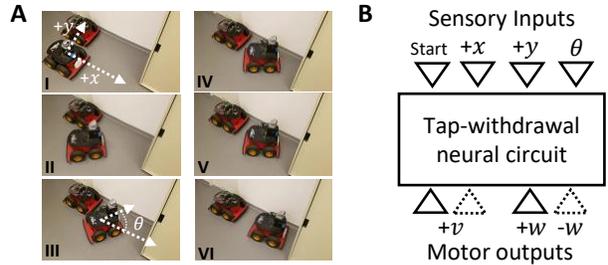

*Figure 4.* Parking setup. A) Environmental setting for the parking task. B) Mapping of the observations and control commands from the environment to the TW circuit.

#### 4.5.1. RL SETTING FOR THE PARKING TASK

A set of checkpoints on a pre-defined parking trajectory were determined in the custom simulated environment. For every checkpoint, a deadline was assigned. At each deadline a reward was given as the negative distance of the rover to the current checkpoint. The checkpoints are placed to resemble a real parking trajectory composed of a sequence of motion primitives: Forward, turn left, forward, turn right, forward and stop. We then learned the TW circuit, by the RL algorithm. The learned policy has been mounted on a Pioneer AT-3 mobile robot and performed a reasonable parking performance. The Video of the performance of the TW neuronal circuit policy on the parking task can be viewed at https://youtu.be/Vwydc2ez9Wc.

## 5. Experimental Evaluation

In this section, we thoroughly assess the results of our experimentation. We qualitatively and quantitatively explain the performance of our neuronal circuit policies. We then benchmark our results where possible, with the existing methods and describe the main attributes of our methodology. We finally illustrate how the activity of the learned neuronal circuits can be interpretable.

### 5.1. Performance

The training algorithm was able to solve all five tasks, after a reasonable number of iterations as shown in Figure 5. All learning curves reached a stable state after the given number of iterations.

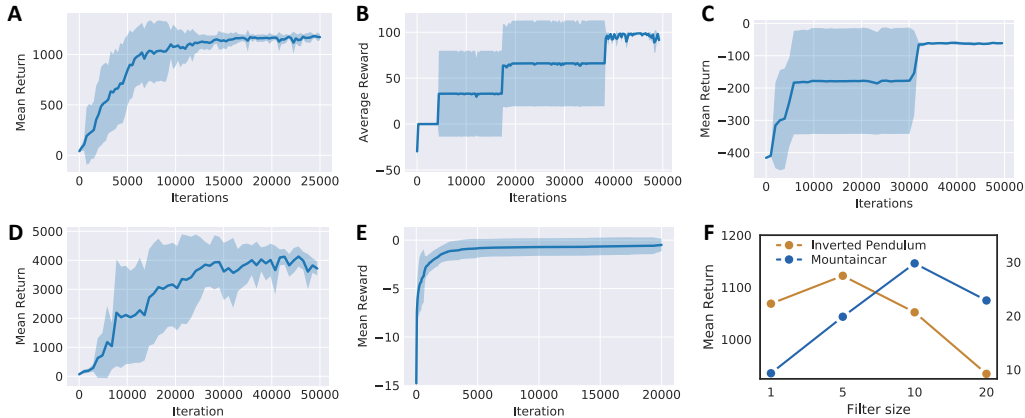

*Figure 5.* Performance as a function of the number of iterations. A) Inverted pendulum B) Mountain car (OpenAI Gym) C) Mountain car (rllab) D) Cart-pole (rllab) E) parking trajectory of the rover robot F) Influence of filter size on training performance

Jumps in the learning curves of the mountain car in OpenAI Gym (Figure 5B) are the consequence of the sparse reward. The inverted pendulum of the Roboschool and the cart-pole of rllab, realized substantially similar learning curves (Figure 5A and 5D) . This indicates the robustness of the policy and the learning algorithm to different frameworks for a given control task. For the deterministic parking trajectory, the learning curve approaches a stable state, exponentially fast.

The final return values for the basic standard RL tasks (provided in Table 2), matches that of conventional policies (Heidrich-Meisner & Igel, 2008), and the state-of-the-art deep neural network policies learned by many RL algorithms (de Froissard de Broissia & Sigaud, 2016; Schulman et al., 2017; Berkenkamp et al., 2017). Table 2, also depicts the average return over the entire training iterations. The average return is not significant compared to the other algorithms applied to the artificial neural network policies (Duan et al., 2016), due to the smaller number of training iterations. Training curves in our case however, reaches an stable state reasonably fast in all tasks even with a fewer number of epochs.

To take advantage of parallel hardware resources and to further increase the performance of our training algorithm, we deployed an ensemble of TW agents which were trained independently in parallel. Depending on the difficulty of the task, between 100% (e.g. Roboschool's inverted pendulum) and 25% (e.g. in Gym's Mountaincar) of the ensembles were able to solve the task within the reported time frame. The values reported in table 2, and Fig5, correspond to the successful cases of the ensemble.

### 5.2. Effect of filter size on the training performance

Fig 2F shows how the choice of the objective-estimate, affects the training performance. The objective-estimate is defined as the mean of the $k$ (=filter size) returns out of 20 rollouts. When $k = 20$, the estimation is equal to the *sample mean*. We tested two environments with different reward settings: The mean return of Mountaincar (OpenAI Gym) after 50,000 training iterations to represent a sparse reward scenario and the mean return of our modified inverted pendulum task after 15,000 training iterations as an example of a gradual reward setting. The results denote that in a sparse reward setting, filtering of the high-outliers tend to degrade the training performance while in a gradual reward setup, filtering of the outliers improves the performance. The reported values in Fig2F, correspond to the average, when running this experiment 10 times. Further discussions about this can be found in the supplementary materials Section 4.

### 5.3. Interpretability of the neuronal circuit policies

Interpretability of neural network policies is still a challenge to be solved (Heess et al., 2016; Bacon et al., 2017). Successful attempts on the development of interpretable representation learning have been provided where the latent space of the learned policies demonstrate interpretable skills (Chen et al., 2016; Florensa et al., 2017). A neuronal circuit policy has the distinct attribute of being interpretable at the cell level. Here, we show how the activity of the learned TW neuronal policies, in different domains is auditable. Figure 6A to 6C represent the normalized membrane potential of neurons in learned policies for the inverted pendulum, the mountain car and the parking task. We describe the global neuronal dynamics in these scenarios.

*Table 2.* Training results. Return value ± standard deviation. Performance of the learned neuronal circuit policies in terms of final return and average return over all training iterations.

| Environment | Final return | Average return |
|---|---|---|
| Inverted pendulum | 1168.5 ± 21.7 | 394.1 ± 80.9 |
| Mountain car (Gym) | 91.5 ± 6.6 | 28.1 ± 9.6 |
| Mountain car (rllab) | −61.3 ± 1.4 | −196.3 ± 78.0 |
| Cart-pole balancing | 3716.1 ± 240.1 | 1230.3 ± 272.2 |
| Parking | −0.49 ± 0.63 | −6.13 ± 1.41 |

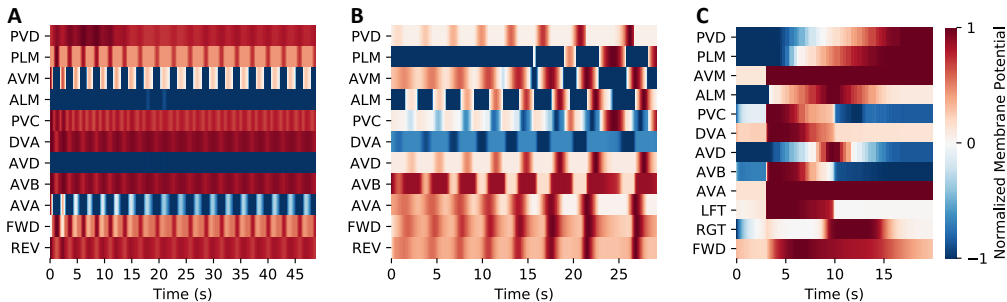

*Figure 6.* Neuronal activity of the Learned neuronal policies. A) The inverted pendulum circuit B) The mountain car (OpenAI Gym) circuit C) The parking circuit. Individual neuron's resting potential, $V_{leak}$, is mapped to 0, neuron's maximum potential is mapped to 1 and neuron's minimum potential is mapped to -1.

**Inverted Pendulum -** Activity of the learned Tap-withdrawal's neurons in a successful episode of the inverted pendulum control, is shown in Figure 6A. The learned circuit surprisingly realized a competitive behavior between two sub-circuits within the network, similar to the reflexive behavior observed in the worm (Wicks et al., 1996). Neurons AVM, AVD, AVA, and REV control the pendulum not to fall on the left side, whereas a circuit composed of PLM, PVC, AVB and FWD, controls the other side of the balance. The antagonistic behavior between the command neurons AVA and AVB balances the pendulum in the middle. Note that DVA neuron acts as a mediator which couples the kinetics of the two sub-circuits. Note that Neurons' dynamics are realized with a variety of time-constants during the balancing control episodes. For instance PVC modulates fast undulations while the motor neurons work with a slower dynamic.

**Mountain Car -** As illustrated in Figure 6B, a gradual increase in the amplitude of the overall periodic dynamic of the neurons is observed due to the alternation of the direction of movement to bring the car uphill. Sensory neurons linearly transform the environmental observations to the global activity of the system, forming a phase-shifted highly synchronized behavior. In every episode, by the rising of PDV (higher negative velocity) or PLM (be at a positive x), all neurons approach their resting potential. This feedback mechanism which prevents unstable behavior, originated from inhibition of PVC and AVD by the ALM sensory neuron. At every period in which the car reaches the maximum position on the left-hand side, a circuit composed of AVM, AVD, AVB and FWD applies a force-pulse which pushes the car to get closer to the target. AVA and AVB retained their objective as in the TW actual circuit, as the command neurons which enforce the network's decision to the motor neurons REV and FWF. However their activity is not fully antagonistic in this learned network. This is where the competition "A more active command neuron, gives rise to a more active downstream motor neuron, and as a results winning the competition", is still present.

**Parking of the rover -** Figure 6C shows the activity of the learned TW policy on the parking task. AVM receives the starting signal and accordingly, a left turn together with moving forward are initiated by the motor neuron FWD and LFT. This is governed by the command neuron AVB. The rover continues moving forward and initializes a right turn by the motor neurons FWD and RGT. This is controlled by AVA. AVD learned to control the turning-phases of the rover by moderating its state, (a higher potential during right turns and a lower membrane state during left turns). PVC and DVA function as the pre-command neurons which incorporate the sensory and the network inputs, into a readable action for the command neuron AVB.

## 6. Conclusions

In this work, we showed that a neuronal circuit policy can be adopted to function as a controller for standard control tasks expressing similar characteristics to the original purpose of the circuit. We illustrated the performance of such policies learned by a search-based RL algorithm, in standard basic RL domains. We also generalized the use of such policies in a real-domain robot control. More importantly, for various domains we showed that our learned policies developed interpretable skills at the neuron level.

Our neuronal policies are limited to the domains in which the sensory observations and motor actions are as many as the available sensory and motor neurons of the neural circuit. Implementation of larger neural circuits to build interpretable controllers for tasks with higher degrees of freedom will be the focus of our continued effort. Moreover, a policy gradient algorithm may significantly enhance the performance of our policies. We open-sourced our policies, to encourage other researchers to further explore the attributes of neuronal circuit policies and apply them to other control and reinforcement learning domains.

## Acknowledgements

Authors would like to thank Jean V. De Carvalho for providing constructive feedbacks on the manuscript. This work was supported with computation resources by Microsoft Azure via the Microsoft Azure for Research Award program.

## References

Ardiel, Evan L and Rankin, Catharine H. An elegant mind: learning and memory in caenorhabditis elegans. *Learning & memory*, 17(4):191–201, 2010.

Bacon, Pierre-Luc, Harb, Jean, and Precup, Doina. The option-critic architecture. In *AAAI*, pp. 1726–1734, 2017.

Bargmann, Cornelia I. Chemosensation in c. elegans. *WormBook*, pp. 1–29, 2006.

Berkenkamp, Felix, Turchetta, Matteo, Schoellig, Angela, and Krause, Andreas. Safe model-based reinforcement learning with stability guarantees. In *Advances in Neural Information Processing Systems 30*, pp. 908–919. Curran Associates, Inc., 2017.

Chalfie, M, Sulston, JE, White, JG, Southgate, E, Thomson, JN, and Brenner, S. The neural circuit for touch sensitivity in *Caenorhabditis elegans*. *Journal of Neuroscience*, 5 (4):956–964, 1985a.

Chalfie, Martin, Sulston, John E, White, JOHN G, Southgate, Eileen, Thomson, J Nicol, and Brenner, Sydney. The neural circuit for touch sensitivity in caenorhabditis elegans. *Journal of Neuroscience*, 5(4):956–964, 1985b.

Chen, Beth L., Hall, David H., and Chklovskii, Dmitri B. Wiring optimization can relate neuronal structure and function. *Proceedings of the National Academy of Sciences of the United States of America*, 103(12):4723–4728, 2006.

Chen, Xi, Duan, Yan, Houthooft, Rein, Schulman, John, Sutskever, Ilya, and Abbeel, Pieter. Infogan: Interpretable representation learning by information maximizing generative adversarial nets. In *Advances in Neural Information Processing Systems*, pp. 2172–2180, 2016.

Culotti, Joseph G and Russell, Richard L. Osmotic avoidance defective mutants of the nematode caenorhabditis elegans. *Genetics*, 90(2):243–256, 1978.

de Froissard de Broissia, Arnaud and Sigaud, Olivier. Actor-critic versus direct policy search: a comparison based on sample complexity. *CoRR*, abs/1606.09152, 2016.

Doya, Kenji. Reinforcement learning in continuous time and space. *Neural computation*, 12(1):219–245, 2000.

Duan, Yan, Chen, Xi, Houthooft, Rein, Schulman, John, and Abbeel, Pieter. Benchmarking deep reinforcement learning for continuous control. In *International Conference on Machine Learning*, pp. 1329–1338, 2016.

Florensa, Carlos, Duan, Yan, and Abbeel, Pieter. Stochastic neural networks for hierarchical reinforcement learning. *arXiv preprint arXiv:1704.03012*, 2017.

Hasani, Ramin M, Beneder, Victoria, Fuchs, Magdalena, Lung, David, and Grosu, Radu. Sim-ce: An advanced simulink platform for studying the brain of caenorhabditis elegans. *arXiv preprint arXiv:1703.06270*, 2017.

Heess, Nicolas, Wayne, Greg, Tassa, Yuval, Lillicrap, Timothy, Riedmiller, Martin, and Silver, David. Learning and transfer of modulated locomotor controllers. *arXiv preprint arXiv:1610.05182*, 2016.

Heidrich-Meisner, Verena and Igel, Christian. Variable metric reinforcement learning methods applied to the noisy mountain car problem. In *European Workshop on Reinforcement Learning*, pp. 136–150. Springer, 2008.

Kato, Saul, Kaplan, Harris S., Schrödel, Tina, Skora, Susanne, Lindsay, Theodore H., Yemini, Eviatar, Lockery, Shawn, and Zimmer, Manuel. Global brain dynamics embed the motor command sequence of *Caenorhabditis elegans*. *Cell*, 163:656–669, October 2015.

Koch, Christof and Segev, Koch. *Methods in Neuronal Modeling - From Ions to Networks*. MIT press, second edition, 6 1998.

Li, Zhaoyu, Li, Yidong, Yi, Yalan, Huang, Wenming, Yang, Song, Niu, Weipin, Zhang, Li, Xu, Zijing, Qu, Anlian, Wu, Zhengxing, and Xu, Tao. Dissecting a central flip-flop circuit that integrates contradictory sensory cues in *C. elegans* feeding regulation. 3:776 EP –, Apr 2012.

Moore, Andrew William. Efficient memory-based learning for robot control. 1990.

Nichols, Annika LA, Eichler, Tomáš, Latham, Richard, and Zimmer, Manuel. A global brain state underlies c. elegans sleep behavior. *Science*, 356(6344):eaam6851, 2017.

Press, William H., Teukolsky, Saul A., Vetterling, William T., and Flannery, Brian P. *Numerical Recipes 3rd Edition: The Art of Scientific Computing*. Cambridge University Press, New York, NY, USA, 3 edition, 2007.

Rankin, Catherine H, Beck, Christine DO, and Chiba, Catherine M. Caenorhabditis elegans: a new model system for the study of learning and memory. *Behavioural brain research*, 37(1):89–92, 1990.

Rastrigin, L. A. About convergence of random search method in extremal control of multi-parameter systems. *Avtomat. i Telemekh.*, 24:1467–1473, 1963.

Russell, Stuart J. and Norvig, Peter. *Artificial Intelligence: A Modern Approach*. Pearson Education, 3 edition, 2010.

Salimans, Tim, Ho, Jonathan, Chen, Xi, and Sutskever, Ilya. Evolution strategies as a scalable alternative to reinforcement learning. 2017.

Schulman, John, Levine, Sergey, Moritz, Philipp, Jordan, Michael I., and Abbeel, Pieter. Trust region policy optimization. *CoRR*, abs/1502.05477, 2015.

Schulman, John, Wolski, Filip, Dhariwal, Prafulla, Radford, Alec, and Klimov, Oleg. Proximal policy optimization algorithms. *arXiv preprint arXiv:1707.06347*, 2017.

Singh, Satinder P and Sutton, Richard S. Reinforcement learning with replacing eligibility traces. *Recent Advances in Reinforcement Learning*, pp. 123–158, 1996.

Spall, James C. *Introduction to Stochastic Search and Optimization*. John Wiley & Sons, Inc., New York, NY, USA, 1 edition, 2003.

Szita, I. and Lörincz, A. Learning tetris using the noisy cross-entropy method. *Neural Computation*, 18(12):2936–2941, 2006. ISSN 0899-7667.

Wen, Quan, Po, Michelle D, Hulme, Elizabeth, Chen, Sway, Liu, Xinyu, Kwok, Sen Wai, Gershow, Marc, Leifer, Andrew M, Butler, Victoria, Fang-Yen, Christopher, et al. Proprioceptive coupling within motor neurons drives c. elegans forward locomotion. *Neuron*, 76(4):750–761, 2012.

White, J. G., Southgate, E., Thomson, J. N., and Brenner, S. The structure of the nervous system of the nematode *Caenorhabditis elegans*. *Philosophical Transactions of the Royal Society of London B: Biological Sciences*, 314(1165):1–340, 1986.

Wicks, SR and Rankin, CH. Integration of mechanosensory stimuli in *Caenorhabditis elegans*. *Journal of Neuroscience*, 15(3):2434–2444, 1995.

Wicks, Stephen R., Roehrig, Chris J., and Rankin, Catharine H. A dynamic network simulation of the nematode tap withdrawal circuit: Predictions concerning synaptic function using behavioral criteria. *Journal of Neuroscience*, 16(12):4017–4031, 1996.

Widrow, Bernard. Pattern recognition and adaptive control. *IEEE Transactions on Applications and Industry*, 83(74):269–277, 1964.

Williams, Ronald J. Simple statistical gradient-following algorithms for connectionist reinforcement learning. *Mach. Learn.*, 8(3-4):229–256, May 1992.

# Supplementary Materials

## 1. Videos of the performance of the learned neuronal circuit policies

| Description | URL |
|---|---|
| TW circuit controls an inverted pendulum at different stages of the training process | https://youtu.be/iOHeQ7DhQv8 |
| TW circuit controls a mountain car at different stages of the training process | https://youtu.be/lMrP1sXp3jk |
| TW circuit performs the parking task | https://youtu.be/Vwydc2ez9Wc |
| Parking task with four degrees-of-freedom | https://youtu.be/jVQqKoHopTU |

Table S1. Videos

## 2. Neural Circuit Implementation and Setup

In this section we describe how to integrate the neuron and synapse equations into a computational framework to build up the TW circuit. Due to non-linearity of the sigmoid function in Eq. (7)Main-text, the neuron's differential equation, Eq. (1)Main-text, becomes non-linear. Unfortunately, there are no complete theory of solving problems of this class, explicitly (Gerald, 2012). Thus, for simulating neural networks composed of such dynamic neuron models, we adopted a numerical implicit solver.

When a network structure is dense and full of synaptic pathways, the set of ODEs (neuron potentials), defined in Eq. (1)Main-text, becomes *stiff* (Press et al., 2007). Therefore, in order to overcome stability issues we used an *implicit* derivation approximation method as follows (Press et al., 2007):

$$\frac{dv}{dt} \approx \frac{v(t) - v(t - \Delta_t)}{\Delta_t} \qquad \text{for some small } \Delta_t. \tag{1}$$

In this way, we discretize the time variable and transform the set of ODEs into a set of iterative equations.

For each neuron, Eq. (1)Main-text, exposed to chemical synaptic currents in the form of Eq. (7)Main-text, and gap junction currents in the form of Eq. (8)Main-text, if we apply approximation of the Eq. (1)supplementary and assume $v_{pre}(t) \approx v_{pre}(t - \Delta_t)$, we can show that the membrane potential of that neuron at time $t$, is computable by:

$$\begin{aligned} v(t) = & \\ & [\frac{C_m}{\Delta_t} v(t - \Delta_t) + G_{Leak} V_{Leak} + \\ & \sum_{i \in Ex} \omega_{ex,i} E_{Rev,ex} + \sum_{i \in Inh} \omega_{inh,i} E_{Rev,inh} + \\ & \sum_{i \in GJ} \omega_{gj,i} v_{pre}(t - \Delta_t)] / \\ & [\frac{C_m}{\Delta_t} + G_{Leak} + \sum_{i \in Ex} \omega_{ex,i} + \\ & \sum_{i \in Inh} \omega_{inh,i} + \sum_{i \in GJ} \omega_{gj,i}] \end{aligned} \tag{2}$$

In Eq. (2)Supplementary, $\omega_{ex,i}$, $\omega_{inh,i}$, $\omega_{gj,i}$, respectively stand for the overall conductance of the excitatory synapse, inhibitory synapse and the gap junction, where $\omega_{ex,i} = g_{ex,i}(v_{pre})$, $\omega_{inh,i} = g_{inh,i}(v_{pre})$, and $\omega_{gj,i} = \hat{\omega}$. Variables together with their boundaries, and constants in Eq. (2)Supplementary, are summarized in the Table 2.

| Parameter | Value | Lower bound | Upper bound |
|---|---|---|---|
| $C_m$ | Variable | 1mF | 1F |
| $G_{Leak}$ | Variable | 50mS | 5S |
| $E_{rev}$ excitatory | 0mV | | |
| $E_{rev}$ inhibitory | -90mV | | |
| $V_{Leak}$ | Variable | -90mV | 0mV |
| $\mu$ | -40mV | | |
| $\sigma$ | Variable | 0.05 | 0.5 |
| $\omega$ | Variable | 0S | 3S |
| $\hat{\omega}$ | Variable | 0S | 3S |

Table S2. Parameters and their bounds of a neural circuit

Formally, Eq. (2)Supplementary, was realized in a hybrid fashion which combine both implicit and explicit Euler's method; The overall neuron equation, Eq. (1)Main-text is approximated by the implicit Euler's method while the parts substituted from Eq. (7)Main-text and Eq. (8)Main-text, were estimated by an explicit Euler's method.

The motivation for implementing such hybrid solver, was to make the resulting algorithm of simulating the neuronal network be separable into the following steps:

1. Compute all the incoming currents, $I_{in}^{(i)}$, form all synapses to the cell using the most recent values of $v(t)$
2. Update all $v(t)$ by Eq. (2)Supplementary

This is significantly effective when implementing a neural network on a real-time controller.

## 3. Experimental Setup Parameters

### 3.1. Experimental setup table

| | Inverted Pendulum | Mountaincar (OpenAI Gym) | Mountaincar (rllab) | Cart-pole | Parking |
|---|---|---|---|---|---|
| Iterations | 25,000 | 50,000 | 50,000 | 50,000 | 20,000 |
| Horizon | 1000 | 1000 | 500 | 500 | 320 |
| Sample size | 20 | 20 | 20 | 20 | 1 |
| Filter size | 10 | 20 | 20 | 10 | 1 |

Table S3. Experiment Parameters

With the aim of gaining a better performance, and utilizing parallel hardware resources and to increase the statistical confidence, all experiments are performed by an ensemble of 12 agents. Due to the stochasticity of the training algorithm, not all agents were able to solve the tasks in the given time frame. The observed success-rates are: Mountaincar (OpenAI gym) 25%, Mountaincar (rllab) 42%, Cart-pole 42%, Inverted Pendulum 100% and Parking 100%.

### 3.2. Neuron's and synapse's parameter-boundaries in the optimization setting

| Type | Lower bound | Upper bound |
|---|---|---|
| $\omega$ | 0 | 3 |
| $\sigma$ | 0.05 | 0.5 |
| $C_m$ | 0.001 | 1 |
| $G_{Leak}$ | 0.05 | 5 |
| $V_{Leak}$ | -90 | 0 |

Table S4. Types of parameters that are optimized and range of valid values

## 3.3. Sensory neuron mappings

As introduced in section 2.2, input and output values are mapped to the potential of sensory respectively motor neurons by an affine mapping. This affine mapping is defined by the minimum and maximum value of the particular input or output value. For each of the five RL environments we set these boundary values separately, according to the following table:

| Environment | Variable | Minimum | Maximum |
|---|---|---|---|
| Inverted pendulum | $x$ | -1 | +1 |
|  | $\varphi$ | -0.12 | +0.12 |
|  | $a$ | -0.3 | +0.3 |
| Mountaincar (OpenAI Gym) | $x$ | -0.02 | +0.02 |
|  | $\dot{x}$ | -0.12 | +0.12 |
|  | $a$ | -1 | +1 |
| Mountaincar (rllab) | $x$ | -0.8 | +0.8 |
|  | $\dot{x}$ | -1.5 | +1.5 |
|  | $a$ | -1 | +1 |
| Cart-pole | $\varphi$ | -0.15 | +0.15 |
|  | $\dot{\varphi}$ | -1 | +1 |
|  | $a$ | -1 | +1 |
| Parking | $x$ |  | +1 |
|  | $y$ |  | +1 |
|  | $\theta$ |  | +1 |
|  | Start signal |  | +1 |
|  | Linear velocity |  | +1 |
|  | Angular velocity | -1 | +1 |

*Table S5.* Input and output boundary values used to define the affine sensory and motor mappings

## 4. Discussion on the influence of the filtering on the training performance

We experienced that a large filter size, i.e an objective estimate, which filters out only few outlying samples, performs better when the reward is sparse. We observed that learning curve of such environments usually has the shape of a series of step functions. One example is the Mountaincar (OpenAI Gym) environment, where a positive reward is only given once at the end, when the entire task has been solved. A large filter size performs better, in such tasks, because the episodes that have been solved by luck, are not filtered out and instead, have a large effect on the estimate. This is crucial for the training in a sparse reward setting, since during the learning phase, we observed that the agent first is unable to solve the task, then is able to solve only a few cases (e.g. when the car starts somewhere uphill) and so on, until the final agent solves the task even in the difficult scenarios.

Furthermore, we observed that a small filter size, i.e. an objective-estimate which is computed only from the lowest samples, performs slightly better in tasks with gradually increasing reward. Learning curve of such environments are usually shaped smoother. An example of such task is the inverted pendulum (Roboschool), where a positive reward is given all the time when the pendulum is still facing upwards. Our filtering strategy performs slightly better here, because, as originally intended, the estimate is not influenced by outlying high samples. In the inverted pendulum task, such samples occur when the pendulum start in an almost straight pose and a high return can be collected without any specific action.

## 5. TW circuit can realize more degrees of freedom

In our first parking experiment, the TW circuit is able to make a turn left and move the rover forward with only one command neuron being active. This means that the circuit is able to solve the task with having binary activation states (active, not active) of the command neuron. To test the flexibility of the TW circuit and underlying neuron model, we set up a second parking experiment. In this experimental setup, we connected the command neuron AVA to two motor neurons responsible for turning right and moving backwards, and AVB to two motor neurons responsible for turning left and moving forward. In this setup, the controller is principally able to move the robot to 4 different directions: Forward, Backward, turn left and turn right. Furthermore, the TW circuit is not able to move the rover forward and turn right with only command neuron being active. If the TW circuit tends to make a right turn and move the rover forward at the same time, (which is necessary to solve this task), the circuit must be able to do this via a synchronized cooperation of the two command neurons. With this configuration, our goal was to test whether the TW circuit can express multiple output primitives with only two command neurons, by operating them in more than two potential states. We conclude that the training algorithm was able to parametrize the TW circuit, such that the agent can keep the trajectory checkpoints, correctly. A video on this scenario can be viewed at https://youtu.be/jVQqKoHopTU.

## 6. Neuron traces of figure 6

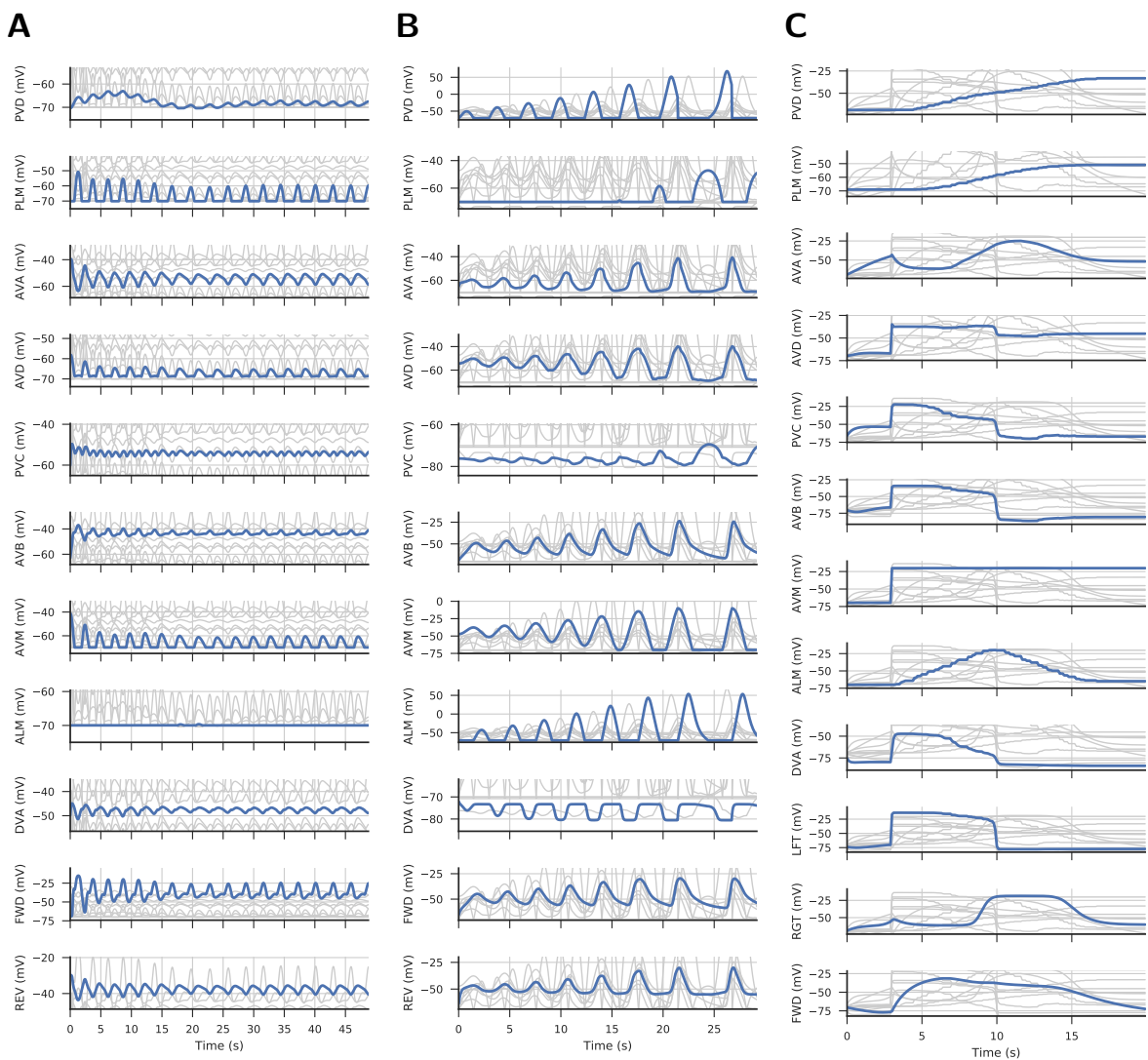

*Figure S1.* Neuron activity represented in figure 6. A) Inverted pendulum, B) Mountain car C) Parking

## References

Gerald, Teschl. *Ordinary Differential Equations and Dynamical Systems*, volume 140 of *Graduate Studies in Mathematics*. American Mathematical Society, 2012.

Press, William H., Teukolsky, Saul A., Vetterling, William T., and Flannery, Brian P. *Numerical Recipes 3rd Edition: The Art of Scientific Computing*. Cambridge University Press, New York, NY, USA, 3 edition, 2007.